\documentclass[10pt,conference]{IEEEtran}
\IEEEoverridecommandlockouts
\usepackage{cite}
\usepackage{amsmath,amssymb,amsfonts}
\usepackage{algorithmic}
\usepackage{graphicx}
\usepackage{textcomp}
\usepackage{xcolor}

\usepackage{enumitem}
\usepackage{url}
\usepackage[T1]{fontenc}
\usepackage[resetlabels]{multibib}
\newcites{T}{Selected Tertiary Studies}
\def\BibTeX{{\rm B\kern-.05em{\sc i\kern-.025em b}\kern-.08em
    T\kern-.1667em\lower.7ex\hbox{E}\kern-.125emX}}

\usepackage{etoolbox}
\makeatletter
\patchcmd{\thebibliography}{%
  \section*{\refname}\@mkboth{\MakeUppercase\refname}{\MakeUppercase\refname}%
}{}{}{}
\begingroup\catcode`#=12
\AtBeginDocument{
  \patchcmd\thebibliography
    {\advance\@tempcnta#1}
    {\advance\@tempcnta#1\else\@tempcnta#1}
    {}{}
}
\endgroup
\makeatother
\begin{document}

\title{Stop Building Castles on a Swamp! The Crisis of Reproducing Automatic Search in Evidence-based Software Engineering 
\thanks{This research is supported in part by Chilean National Research and Development Agency (ANID, Chile) under Grant
FONDECYT Iniciaci{\'o}n 11180905.}
}

\author{\IEEEauthorblockN{Zheng Li}
\IEEEauthorblockA{\textit{Department of Computer Science} \\
\textit{University of Concepci{\'o}n}\\
Concepci{\'o}n, Chile \\
ORCID: 0000-0002-9704-7651}
}

\maketitle

\begin{abstract}
The evidence-based approach has increasingly been employed to synthesize empirical findings from the primary research in software engineering. Nevertheless, the \textit{reproducibility} of evidence-based software engineering (EBSE) studies seems to be underemphasized. In our investigation into the automatic search of 311 sample studies, more than 50\% of the search strings are not reusable; about 87.5\% of the search activities (e.g., search field settings) are unrepeatable; and more than 95\% of the whole automatic search implementations are unreproducible. Considering that searching is a cornerstone of an EBSE study, we are afraid that the reproducibility of the current secondary research could be worse than we can imagine. By analyzing and reporting the root causes of the aforementioned observations, we urge collaboration and cooperation among all the stakeholders in our community to improve the research reproducibility in EBSE.

\end{abstract}

\begin{IEEEkeywords}
automatic search, digital libraries, EBSE, reproduction crisis, systematic literature review
\end{IEEEkeywords}

\section{Introduction}
Given the fast development of evidence-based software engineering (EBSE), secondary studies driven by EBSE's main methodology (namely systematic literature review or SLR) \cite{Zhou_Jin_2016} have increasingly been conducted and reported. However, although various SLR guidelines all emphasize the rigorousness and reproducibility of evidence-based research \cite{Kitchenham_Charters_2007,Mariano_2017,Okoli_2015}, \textit{reproduction} seems to be insufficiently discussed in the EBSE community and frequently confused with the role of replication. 
According to the consensus across multiple scientific disciplines, a ``research is reproducible if another researcher actually uses the available data and code and obtains the same results'' \cite{NASEM_2019}. As such, being reproducible is a necessary condition for a scientific study to be considered sound and valid \cite{Miyakawa_2020}, while only valid studies are meaningful (or even possible) to be replicated to confirm and reinforce the empirical findings \cite{Shepperd_Ajienka_2018}. Considering that SLR could suffer from a wide range of validity threats \cite{Zhou_Jin_2016}, we are particularly concerned with the reproducibility of the existing SLR implementations in the software engineering domain.

Our current work in progress is focused on Searching that has been treated as the most crucial stage of the whole SLR process \cite{Bailey_Zhang_2007}. It is known that paper selection (based on searching) could involve significant biases  \cite{Zhou_Jin_2016}. 
 Several reproduction studies in EBSE even claim that it is barely possible to select the same set of papers to review, unless the studies are conducted by experts in a relatively narrow area \cite{Wohlin_Runeson_2013}. Thus, before selecting papers, it is worth concentrating on the reproducibility of searching, in order to narrow down the root causes of biased paper selection. Since automatic search generally has less human interferences than manual search, we conservatively define our research question as:


\begin{enumerate}[leftmargin=*,labelindent=10pt,label={\textbf{RQ:}}]
    \item To what extent can we reproduce the automatic search of the existing SLR implementations in EBSE? 
\end{enumerate}

To facilitate answering this research question, we further define three indicators and the corresponding criteria to measure the reproducibility of automatic search, as specified in Section \ref{sec:criteria}. Ideally, if the automatic search is largely reproducible, we will at least have a solid common ground to investigate the biases in the other SLR stages. Unfortunately and surprisingly, after retrying the automatic search of 311 sample studies (cf.~Section \ref{sec:sampling}) against their original reports, we see that more than half of the search strings are not reusable; about seven eighths of the search activities (e.g., search field settings) are unrepeatable; and more than 95\% of the whole automatic search implementations are unreproducible. Accordingly, we have to estimate that the reproducibility of the published EBSE studies is generally worse than we can imagine.

By respectively analyzing the causes of the aforementioned observations, we have identified not only the researchers' mistakes but also the search engine limits, as detailed in Section \ref{sec:results}. For example, the key information (e.g., the search string) for reproduction might be missing in the SLR papers; while the recent changes in some search sources could make previously valid automatic search unreproducible (e.g., the reported search string cannot be inputted successfully now). These identified causes can in turn act as a checklist to remind researchers of the existing issues and limits for conducting EBSE studies in the future. 

More importantly, our investigation reveals that improving the research reproducibility in EBSE requires collaboration and cooperation within the whole community, as discussed in Section \ref{sec:conclusions}. In general, we argue that reproducibility-oriented reporting regulations are urgently needed for the ease of reference by both researchers and peer reviewers. When focusing on the searching only, since search engine limits are out of control of the researchers, there are urgent needs to standardize the interfaces/engines of academic search sources.

\section{Three Indicators and Measurement Criteria}
\label{sec:criteria}
Recall that a reproducible research allows another researcher to use the available data and code to obtains the same results \cite{NASEM_2019}. We regard search string(s), search field(s), and time period as the critical data and code for conducting and reproducing an automatic search, because modifying or missing any of them would lead to significantly different results. The search results here are directly represented by the amount of hits of publications, for the purpose of convenient comparison. 
Correspondingly, we employ three indicators and define their respective criteria to help verify and measure the reproducibility of previous automatic search practices, as specified below.

\subsubsection{\textbf{Reusability of Search Strings}} 
Being derived from research questions, search strings play a fundamental role in driving the automatic search in a specific study. Therefore, we treat search strings as the most crucial data against the others. By relaxing the settings of search fields and time period as well as relaxing the concern about search results, we adopt this indicator to verify whether or not the previously reported search strings can still be reused today. We define a search string to be reusable as long as it can be processed in the default or any flexible console (e.g., the command search window) of the reported search sources. 

\subsubsection{\textbf{Repeatability of Search Activities}}
Before checking results, a study can already be unreproducible when its available data and code become invalid as time goes by. Therefore, we propose to further validate the search strings together with the search fields and time period from a previous study, by trying only to repeat the study's search activities while ignoring the results. We define the automatic search activities of a study to be repeatable when and only when the reported search strings can be processed in the reported search sources, with exactly the same settings of search fields and time period.

\subsubsection{\textbf{Reproducibility of Automatic Search}} 
Given the aforementioned clarification, we finally define an automatic search to be reproducible when and only when the search activities can not only be repeated by following the reported instructions/settings, but also bring the same (or at least close) results from the reported search sources. Note that, considering the possible indexing delays in the digital libraries, we have relaxed the condition of comparing search results. In specific, we regard the new results as close to the reported results, as long as they have less than an order of magnitude difference.

\section{Sampling EBSE Studies}
\label{sec:sampling}
Since introduced by Kitchenham in 2004 \cite{Kitchenham_Charters_2007}, the methodology SLR has been widely employed not only in the software engineering domain but also in the broad areas of computer science and engineering. As a result, it is difficult to gather the population of pure EBSE studies. Therefore, we resort to the \textit{Snowball Sampling} technique to scope a representative subset of the population, by choosing relevant tertiary studies as the initial ``snowballs''. A tertiary study is a study that follows the same methodology SLR to summarize or synthesize the existing secondary research to answer wider research questions \cite{Kitchenham_Charters_2007}. Thus, we can conveniently retrieve a relatively large amount of secondary studies from even a small set of tertiary ones. More importantly, we can meanwhile guarantee the samples' relevance, as they have been carefully filtered into the context of the selected tertiary studies (EBSE in this case).

\subsection{Searching Tertiary Studies}
Given the nature of sampling, it is not necessary to pursue the completeness of tertiary study selection, either. Therefore, we decide to rely on a single search source rather than employ the multi-source strategy. Since Google Scholar has been proven to have a considerably broad coverage of scientific publications \cite{Wohlin_Runeson_2013}, we use Google Scholar to compensate the indexing bias of the single-source search to some extent. Moreover, we follow the short-string strategy \cite{Li_Avgeriou_2015} to try to maximize the search scope:

\texttt{(\textquotedbl tertiary study\textquotedbl ~software review)} 

According to our pilot trials for building this search string, the keyword \texttt{software} restricts tertiary studies to be software-related at least, while the keyword \texttt{review} helps filter out a large amount of studies on tertiary education. Note that the conjunction operator AND is omitted intentionally, otherwise \texttt{AND} could also be treated as a search term by Google (cf.~Section \ref{subsubsec:limits4repeat}).

The search was conducted on May 28, 2020.
We only searched the tertiary studies published in the past five years (from 2015 to present, i.e.~to May 28, 2020), as they must have reviewed more secondary studies than before since the ten-year application of SLR in software engineering. After unchecking ``include patents'' and ``include citations'', Google Scholar returned ``about 2160 results''.

\subsection{Selecting Relevant Tertiary Studies}
To improve the selection efficiency, we rely on Google's ``sort by relevance'' to limit our screening to the first 20 pages of the search results (10 results per page). 
 The decision about the 20-page threshold is based on the similar rule \cite{Williams_Carver_2010} that we stop screening when we do not find any tertiary study on software engineering topics for pages, by reading the result titles and snippets.

When it comes to selecting individual tertiary studies, we further regulate to only include peer-reviewed journal and conference papers in which their identified secondary studies are specified. In other words, we exclude those tertiary studies that are lacking in quality evaluation or impossible for snowball sampling. Eventually, 11 relevant tertiary studies \citeT{Barros-Justo_2019,Budgen_Brereton_2018,Curcio_Santana_2019,Goulao_Amaral_2016,Hoda_2017,Khan_2017,Marimuthu_Chandrasekaran_2017,Nurdiani_2016,Raatikainen_2019,Rios_Neto_2018,Zhou_Zhang_2015}  were selected for us to collect EBSE research samples.

\subsection{Collecting Samples of EBSE Research}
By referring to the discussions and references of the 11 tertiary studies, we extracted 491 records of secondary studies to act as candidate samples of EBSE research. However, not all the candidates are valid samples in our work. In addition to removing duplicates, we also removed one paper written in Spanish, six non-SLR survey papers, and seven SLR papers whose search strategy is purely manual.  At last, 311 studies are involved in our investigation into the reproducibility of automatic search in EBSE.

\section{Results of Reproducibility Analysis}
\label{sec:results}
\begin{figure}[htbp]
\centerline{\includegraphics{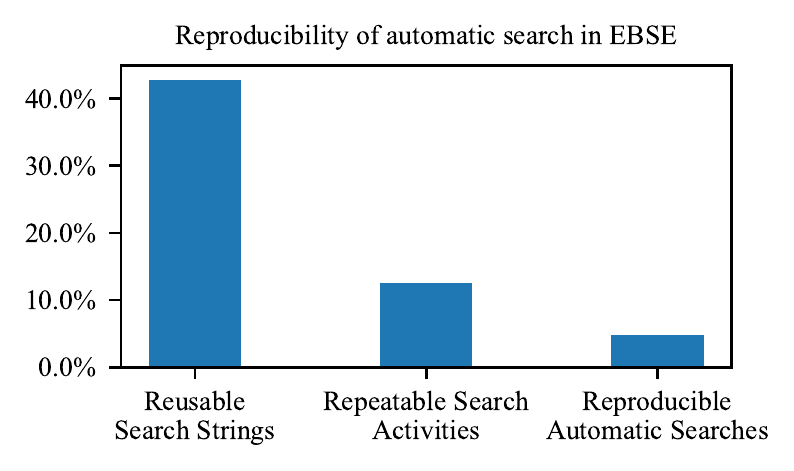}}
\caption{Reproducibility analysis results measured by the three indicators.}
\label{fig}
\end{figure}

After verifying the automatic search of every single study against the predefined criteria, we obtained three indicator ratios, as illustrated in Fig.~\ref{fig}. In detail, although most studies have reported search strings, only about 43\% of the samples (133 out of the 311 studies) can pass our reusable-string verification. By including search fields and time period, we can repeat the search activities described in about one eighth of the samples (39 studies). By further comparing the results, surprisingly, we can only reproduce the automatic search for less than 5\% of the samples (15 studies), even after including two studies with fixable syntax errors.

During the verification, we particularly analyzed the causes of being unable to reproduce the original work.\footnote{The verification and analysis details are shared at \url{http://doi.org/10.5281/zenodo.4447489}} 
 The analysis results are classified and summarized with respect to the three indicators, as described in the following subsections.

\subsection{Causes of Non-Reusable Search Strings}
It is known that the lack of standardization between search engines would limit the reusability of search strings even in the same study \cite{Jabangwe_2015}. To address this limit, 14 studies clearly published alternative strings for individual search sources (e.g., \cite{Marques_2019}). Although many papers claim to have constructed ``semantically and logically equivalent'' strings, we have to treat this as a perfunctory statement due to the untraceable information. 
Besides this typical problem, we distinguish between the reporting issues and engine limits that make search strings non-reusable.    

\subsubsection{The Reporting Issues}
We have observed both fixable and unfixable reporting issues.

The fixable issues are generally typos or syntax errors in the reported search strings, such as, boolean operators are misspelled \cite{Kasoju_Petersen_2013}, duplicated \cite{El-Attar_2012}, or mixed with non-standard symbols (e.g., \cite{Bashroush_Garba_2017}); 
 quotation marks are missing (e.g., \cite{Afzal_Mahmood_2016}) or unpaired \cite{Ahnassay_Bagheri_2013}; opening and closing parentheses do not match (e.g., \cite{Stavru_Krasteva_2012}). 
 A hidden problem here is that, since many search engines do not check syntax (not to mention typos), copying the incorrect strings can still be processed and will eventually flaw the automatic search imperceptibly.

As for the unfixable issues, the extreme case is that the search strings are completely missing. In particular, six studies did not mention search strings at all in their publications. 
 In another five papers, readers are referred to external materials (technical reports or online attachments) for details, while the external materials are unavailable at the time of our investigation. 
 A special case is that one of the two strings used in \cite{Galindo_Benavides_2019} was claimed to be from an earlier study \cite{Benavides_Segura_2010}; however, the earlier study did not report any search string. 

Although some other studies elaborated more about search strategy, a frequent unfixable issue of theirs is the insufficient information for rebuilding search strings. For instance, four studies elaborately explained the rules of combining search terms, while the search terms were not (fully) specified. 
 On the contrary, 17 studies exhaustively listed the search terms, while no specification was given about how to combine them. 
 Even when the rules and search terms are both available, we are unable to reproduce the original search strings of ten studies, mainly due to the combination explosion. 
 For example, the authors claimed to build strings by taking all the combinations of one item from a 29-term list and one or more items from another 44-term list \cite{Lopez-Herrejon_2015}; given such descriptions, it is still barely possible for readers to know how many and what search strings were used originally.

\subsubsection{Search Engine Limits}
It is evident that the search engines of all the popular digital libraries have been evolving. For example, ACM DL that used to have string length limit can work well with long search strings now. Nevertheless, the evolution does not always bring beneficial features to users. For example, ScienceDirect seems to have more restraints than before. Consequently, some previously well-designed search strings are not reusable any more. 

In brief, the main engine limits we have met include: 
\begin{itemize}
\item ACM DL does not support the proximity operators.
\item Google Scholar has a maximum limit of 256 characters.
\item IEEE Xplore has a maximum limit of six wildcards.
\item ScienceDirect does not support wildcard search.
\item ScienceDirect has a maximum limit of 500 characters.
\item ScienceDirect has a maximum limit of eight boolean operators per search field.
\end{itemize}

In addition, when verifying the search string of \cite{Paternoster_Giardino_2014}, we receive an alert from Web of Science (WoS) saying that three or more characters are required before a truncation symbol (due to the term \texttt{software start up*}). We believe that IEEE Xplore also has this requirement, although the term \texttt{us*} used in \cite{Senapathi_Srinivasan_2013} triggers a failure response of ``400 Bad Request'' instead of any meaningful message.

\subsection{Causes of Unrepeatable Search Activities}
Similarly, here we also distinguish between the reporting issues and search engine limits.
\subsubsection{The Reporting Issues}
As the worst case, six studies did not mention any search source, even though they all reported (at least partial) search strings. 
 
Among the reported search sources, we see some retired ones and unknown ones (e.g., Scirus and Computer Database respectively). In this work we intentionally ignore these two situations from the repeatability analysis, because it makes no sense to argue about the former, and we treat the latter as the limit of our knowledge.

For the available sources, surprisingly, 204 studies did not specify either the search fields or the time period. Note that the declared ``no restriction'' of fields or time has been counted as a given specification.

\subsubsection{Search Engine Limits}
\label{subsubsec:limits4repeat}
Although many studies highlighted the diversity in the search source interfaces, few of them  precisely described field settings with regard to individual engines. Therefore, we have conducted conservative analysis by tolerating some trivial differences between the original reports and our physical trials. For example, even when the full-text search is not specifically supported by an engine (e.g., WoS), we still consider it repeatable as long as the all-field (or anywhere) search is available.

As for the major engine limits, to save space, we only list a set of common ones that prevent us from repeating search activities of multiple studies:
\begin{itemize}
\item ACM DL has removed the search field of Review. 
\item Google Scholar and ScienceDirect do not have the filter of Subject Area. 
\item Google Scholar and SpringerLink do not support field-specific search, except for the field of Title.
\item Google Scholar does not recognize wildcards and the boolean operator AND in its title search.
\end{itemize}

In particular, IEEE Xplore does not allow the search string's (or term's) first character to be an opening parenthesis in any specific field, which is a hidden, inconvenient and misleading restraint. If violated, it will merely return ``No results found''. Although the command search without specifying fields can work with parenthesized strings, the functional effect is only equivalent to the all-metadata search by default. Thus, when applying sophisticated strings to non-metadata search in IEEE Xplore, the search fields must be manually and repeatedly specified inside of each pair of parentheses.

\subsection{Causes of Unreproducible Automatic Search}
In fact, the previously identified causes have already made the reported automatic searches unreproducible. To avoid duplication, we clarify that the analysis here only includes new causes that have not been revealed by the previous investigations into search strings and activities. 

The unique step of this analysis is to verify the search output. Among those having at least search strings as input, 120 studies did not specify the results from individual search sources, which makes their output verification impossible. As for the others, many reported results are significantly different to that obtained in our trials.  For example, by applying an easy-to-reuse string to IEEE Xplore \cite{Ribeiro_Farias_2016}, the original study collected 234 papers published till 2014 via the title and abstract search, whereas we can receive only 21 hits even by the all-metadata search (one hit if title search and nine hits if abstract search). Apparently there must be something wrong, while unfortunately it is hard to know if this is caused by any reporting issue or by later changes in the digital library.

At last, a special observation is that five studies described inconsistent search sources in different places of their research papers. 
For instance, some digital libraries listed in the search strategy part are missing or replaced with some others in the search implementation part. Such an inconsistency makes those studies not only unreproducible but also unconvincing.

\section{Conclusions}
\label{sec:conclusions}

In this research, we can only reproduce the automatic search of less than 5\% of the 311 sample SLR papers. Considering that searching is an early-stage cornerstone for any SLR implementation, we are afraid that the validity of many published EBSE studies could have been flawed by their fundamentally poor reproducibility. Driven by this conclusion, we propose two suggestions to address (or at least relieve) the potential reproducibility crisis in EBSE.

Firstly, although the implementation guidelines of SLR have been well documented, we urge the community to define reproducibility-oriented regulations for reporting and reviewing EBSE studies. 
It is noteworthy that resorting to external materials to share key information does not seem to work. In our investigation, among the nine studies that referred readers to external materials for either search strings, settings, or results, only one study's technical report is still available \cite{Galster_2014}. In an unusual case, the researchers developed a website to document all the details in order to make their EBSE study ``repeatable for other researchers and research groups'' \cite{Neto_2011}. 
 Nevertheless, such expectation does not come true, because the website now has nothing left but only the homepage.

Secondly, we urge all the stakeholders in the community to take responsibility for improving the research reproducibility in EBSE (e.g., setting up of an independent committee on standardizing the popular digital libraries), rather than purely emphasize the obligation of, and rely on, the researchers. In the context of automatic search, many researchers have met and reported troubles with the variations in the search mechanisms of different engines, while the new changes in some digital libraries seem to have made the situation even worse \cite{Elsevier_2020}. Therefore, we particularly argue that even though different libraries might have business competitions, the competition barriers should be their indexed and stored publications instead of the reinvented wheels of search engines. 

~\nociteT{Barros-Justo_2019,Budgen_Brereton_2018,Curcio_Santana_2019,Goulao_Amaral_2016,Hoda_2017,Khan_2017,Marimuthu_Chandrasekaran_2017,Nurdiani_2016,Raatikainen_2019,Rios_Neto_2018,Zhou_Zhang_2015}

\bibliographystyle{IEEEtran}
\bibliography{NIER}

\makeatletter
\renewcommand\@bibitem[1]{\item\if@filesw \immediate\write\@auxout
    {\string\bibcite{#1}{T\the\value{\@listctr}}}\fi\ignorespaces}
\def\@biblabel#1{[T#1]}
\makeatother
\bibliographystyleT{IEEEtran}
\bibliographyT{TS}

\end{document}